\documentclass[showpacs,preprintnumbers,amsmath,amssymb,twocolumn,nofootinbib,nobibnotes,longbibliography]{revtex4-2}
\usepackage{color}
\usepackage[unicode=true,colorlinks=true,citecolor=blue]{hyperref}
\usepackage{epstopdf}
\usepackage{graphicx}
\usepackage[normalem]{ulem}

\usepackage[english]{babel}
\usepackage{tabularx}

\usepackage{dcolumn}
\usepackage{multirow}
\usepackage{bm}
\usepackage{upgreek}
\usepackage{physics}

\newcommand{\St}[1]{\mathop{\rm St}{#1}}

\usepackage[T2A]{fontenc}
\usepackage[cp1251]{inputenc}

\begin{document}

\title{
Spin and Valley Hall effects induced by asymmetric interparticle scattering
}

\author{M.~M.~Glazov}
\author{L.~E.~Golub} 	
\affiliation{Ioffe Institute,  	194021 St.~Petersburg, Russia}


\begin{abstract}
We develop the theory of the spin and valley Hall effects in two-dimensional systems caused by asymmetric -- skew -- scattering of the quasiparticles. The collision integral is derived in the third order in the particle-particle interaction with account for the spin-orbit coupling both for bosons and fermions. It is shown that the scattering asymmetry appears only in the processes where the interaction between the particles in the initial and intermediate state is present. 
We show that for degenerate electrons or nondegenerate particles the spin and valley currents induced by interparticle collisions are suppressed  with their steady-state values being proportional to the squared temperature or density, respectively. Our results imply non-Fermi liquid properties of electrons in the presence of electron-electron skew scattering. 
Strong deviations from conventional picture of interparticle scattering are also demonstrated for the skew scattering of two-dimensional degenerate bosons, e.g. excitons or exciton polaritons: The spin or valley current of degenerate bosons contains the enhancement factor exponentially growing with increase in the particle density. 
\end{abstract}

\maketitle

\section{Introduction}

Spin-dependent effects in condensed matter result from a coupling between the electron orbital and spin degrees of freedom caused by the spin-orbit interaction. Spin currents~\cite{Rashba:SHE:review} take a special place among of the most exciting phenomena in the field both from fundamental and applied physics viewpoints.  
An interesting issue here 
is the possibility to have a spin current in thermodynamic equilibrium with the most prominent example being the Rashba medium -- a system with linear in the momentum spin-dependent terms in the electron or hole Hamiltonian~\cite{rashbasheka,rashba64,Rashba1988175} -- where the spin can flow even if the system is not perturbed~\cite{rashba:241315,PhysRevB.76.033306}. However, in a uniform media these currents do not result in any spin accumulation. The spin transport and spin accumulation arises in the presence of an external electric field where the spin Hall effect (SHE) arises~\cite{dyakonov71,hirsch99}, see Ref.~\cite{Rashba:SHE:review} for review. It results in a conversion of the particle flux to the transverse spin current and subsequent accumulation of the spin polarization at the sample's edges~\cite{dyakonov_book}. Generalization of this phenomenon to the multivalley systems is the valley Hall effect (VHE), i.e., the  conversion of the particle flux to the perpendicular opposite flows of particles belonging to two different valleys~\cite{Mak27062014,2020arXiv200405091G, Glazov2020b,Glazov_2021b}.

The microscopic mechanisms of SHE and VHE are naturally related to the spin-orbit coupling and can be both intrinsic and extrinsic~\cite{engel:166605}, and the dominant one in many cases is the skew-scattering mechanism associated with the spin- or valley-dependent asymmetry of the particle scattering by a defect~\cite{Mott425}. 
This skew scattering has been studied in both bulk semiconductors and heterostructures as well as in metals for the scattering by both impurities and acoustic phonons~\cite{gy61,abakumov72,2020arXiv200405091G}. The possibility of an asymmetric -- skew -- electron-electron scattering has been briefly discussed in the literature in the context of the spin Hall drag in double quantum well structures~\cite{badalyan-2009} and recently in Ref.~\cite{Glazov_2021b} for the spin accumulation in the hydrodynamic electron transport regime, see also Ref.~\cite{fruchart2022odd}. However, the theory of the electron-electron skew scattering is far from being complete.

In this work, we consider the SHE and VHE caused by the inter-particle collisions only. This situation is highly relevant, e.g. for two-dimensional (2D) high-mobility semiconductors and graphene where electron-electron scattering dominates over the impurity and phonon scattering and controls the transport effects~\cite{PhysRevB.51.13389,Bandurin1055,Moll1061,Sulpizio:2019uc,Gusev:2020vd,Ku:2020ue,Ratchet_hydro,PhysRevLett.126.076803}, see Ref.~\cite{Narozhny:2022ud} for review. 
We argue that the skew scattering and, hence, the SHE and VHE are possible at the inter-particle scattering. 
We derive the kinetic equation with allowance for the skew scattering processes both for fermions and bosons and demonstrate asymmetric in spin and momenta terms in the collision integral. They appear beyond the Born approximation for the scattering amplitude. We show that for the spin or valley Hall effect to take place, the scattering should occur due to the interaction of the particles in the initial and intermediate states.
It results in 
unconventional behavior of SHE and VHE in both degenerate Fermi and Bose as well as in nondegenerate gases.

The paper is organized as follows. In Sec.~\ref{sec:kinetic} we present the kinetic equation and derive the collision integral with allowance for the asymmetric scattering in the third-order perturbation theory. We demonstrate derivation by two methods, using the scattering matrix and using the Keldysh diagram technique. Section~\ref{sec:she} contains the theory of the SHE and VHE at the interparticle collisions and particular results for the Fermi, Boltzmann and Bose systems. 
General implications of the obtained results and conclusion are presented in the end of the paper, Sec.~\ref{sec:concl}. Details of calculations are given in Appendices.

\section{Interparticle skew collision integral}
\label{sec:kinetic}

We consider a two-dimensional system in the $(xy)$ plane and assume that the quasiparticles are characterized, in addition to the wavevector $\bm k$, by a spin or valley index $s=\pm$. In conventional semiconductor quantum wells with the two-dimensional electron gas $s$ is the $z$ component of the electron spin, while in the transition-metal dichalcogenide monolayers $s$ distinguishes two valleys $\bm K_+$ ($s=+$) and $\bm K_-$ ($s=-$)~\cite{2020arXiv200405091G,2053-1583-2-2-022001,Glazov_2021b}. In these systems, the two valleys are related by the time-reversal symmetry and share the same properties with the spin component. In the case of Bose quasiparticles, for instance, excitons in quantum wells or transition-metal dichalcogenide monolayers or exciton-polaritons in microcavities, $s=\pm$ denote the polarization of exciton ($\sigma^+$ or $\sigma^-$) also known as the exciton pseudospin and some times termed as a valley index in two-dimensional transition-metal dichalcogenides~\cite{ivchenko05a,kavokin05prl,glazov2014exciton,PhysRevLett.115.166804,Glazov2020b,Lundt:2019aa}.

In standard conditions of the SHE or VHE the quasiparticles are driven by an external force, real electric field in the case of electrons or synthetic fields for excitons and exciton-polaritons~\cite{2020arXiv200405091G,Glazov2020b,Gianfrate:2020aa} and the spin- or valley current is detected in the transversal direction. Hereafter we take into account the spin- and valley-dependent contributions to the scattering matrix elements that are related to the $\bm k\cdot\bm p$-mixing with the remote bands, see Refs.~\cite{2020arXiv200405091G,Glazov2020b,Glazov_2021b} for details. Moreover, we focus on the asymmetric or skew scattering contributions to the SHE and VHE which can be considered independently of the anomalous contributions, side-jump and anomalous velocity. The side-jump effect in the case of electron-electron scattering has been studied in detail in Refs.~\cite{PhysRevLett.121.226601,Glazov_2021b}. Thus, we can describe SHE and VHE within the kinetic equation approach and introduce the distribution functions $f_{\bm k,\pm}$ for the quasiparticles with a given spin or valley index $s$, while off-diagonal in $s$ elements of the density matrix are unimportant for the present study.

\begin{figure*}[ht]
\includegraphics[width=0.75\textwidth]{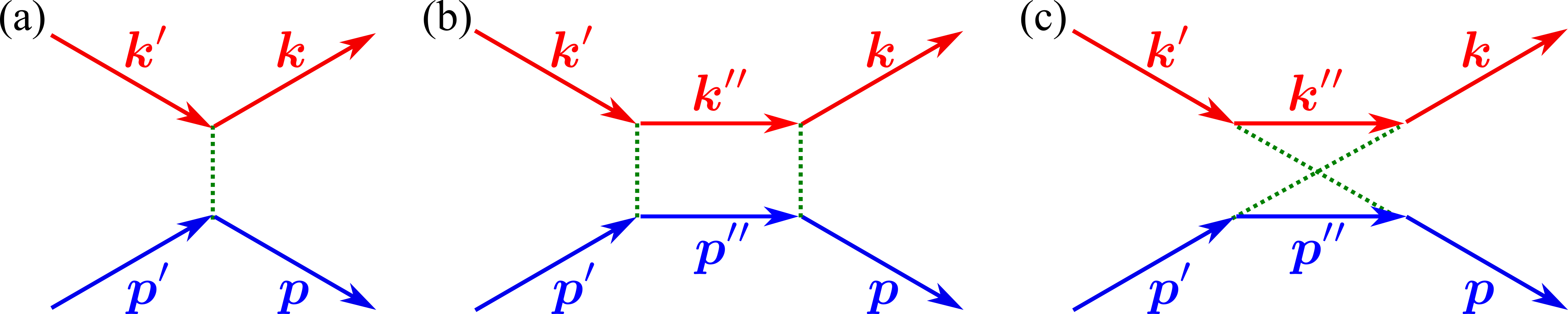}
\caption{Sketch of the scattering processes contributing to the skew effect. (a) First-order contribution. (b) and (c) Second-order contributions. Red and blue arrows denote the electrons with $s=+$ and $-$, respectively, green dotted line shows the interaction.
} 
\label{fig:scatt3}
\end{figure*}

We present the steady-state kinetic equation for the distribution functions $f_{\bm k,s}$ in the following form
\begin{equation}
\label{kinetic:gen}
\frac{\hbar}{m}(\bm F \cdot \bm k) 
\dv{f^{0}_k}{\varepsilon_k}
+ \frac{\delta f_{\bm k,s}}{\tau_p} = \St[f_{\bm k,s}].
\end{equation}
Here $\bm F$ is the force acting on the quasiparticles due to the real or synthetic electric field, $f^{0}_k$ is the equilibrium distribution function,
$\delta f_{\bm k,s}= f_{\bm k,s}- \langle f_{\bm k,s} \rangle$ is the anisotropic correction and the angular brackets denote the angular averaging,
the energy $\varepsilon_k=\hbar^2k^2/(2m)$ with
$m$ being the effective mass of the quasiparticle, $\tau_p$ is the momentum relaxation time caused by the  scattering off the static disorder or phonons,\footnote{We assume that the disorder is Gaussian and neglect two-phonon processes, in this case the disorder and phonon scattering do not have asymmetric component.} and $\St[f_{\bm k,s}]$ is the interparticle collision integral. It contains the symmetric and antisymmetric parts. The symmetric part of the collision integral is responsible for thermalization of quasiparticles, their viscosity and, importantly, relaxation of the spin or valley current~\cite{glazov02,amico02,amico:045307,glazov04a,weber05}.  The asymmetric part is responsible for the spin or valley Hall effect.

The scattering matrix element ${\bm k',s_1;\bm p', s_1'} \to {\bm k,s;\bm p,s'}$ by a (long-range) interaction  potential $V(\bm r_1-\bm r_2)$ reads
\begin{multline}
\label{me:as}
\left<\bm k, s; \bm p, s'\right| V \left|\bm k', s_1; \bm p', s_1' \right> = \delta_{\bm k+ \bm p, \bm k'+ \bm p'}\\
\times  \left[V(|\bm k - \bm k'|)\langle {\bm k,s}|{\bm k',s_1}\rangle\langle {\bm p,s'}|{\bm p',s_1'}\rangle \right. \\
\pm 
\left. (\bm k' \leftrightarrow \bm p', s_1 \leftrightarrow s_1')\right].
\end{multline}
Here the top sign ($+$) refers to the bosons and the bottom sign ($-$) refers to the fermions, we introduced the 2D Fourier-image of the interaction potential, $V(q)$, and used the bra-ket notation $\langle {\bm k,s}|{\bm k',s_1}\rangle$ for the overlap integral of the Bloch amplitudes $v_0^{-1}\int_{v_0} u_{\bm k,s}^*(\bm \rho)u_{\bm k',s_1}(\bm \rho) \dd\bm \rho$ where the integration is performed over the unit cell volume $v_0$.

Here and in what follows we consider collisions of particles (bosons or fermions) with different spin or valley
indices $s$ and $s' \neq s$. The collisions of the particles with
the same $s = s'$ do not produce any spin or valley current owing to the momentum conservation, and do not
significantly contribute to the relaxation of the spin and
valley currents~\cite{Glazov_2021b}; and we disregard spin- or valley-flip processes. Taking $s'=-s$ in Eq.~\eqref{me:as} we have:
\begin{multline}
\label{me:as_1}
\left<\bm k, s; \bm p, -s\right| V \left|\bm k', s_1; \bm p', s_1' \right> = \delta_{\bm k+ \bm p, \bm k'+ \bm p'}\\
\times  \left[V(|\bm k - \bm k'|)\langle {\bm k,s}|{\bm k',s}\rangle\langle {\bm p,-s}|{\bm p',-s}\rangle \right. \delta_{s_1,s}\delta_{s_1',-s} \\
\pm 
\left. (\bm k' \leftrightarrow \bm p', s_1 \leftrightarrow s_1')\right].
\end{multline} 
We see that the matrix element in Eq.~\eqref{me:as_1} is non-zero only if
(i) $s_1 = s$ and $s_1' = -s$ or (ii) $s_1' = s$ and $s_1 = -s$.
Unlike interparticle scattering
with the same spins or valley indices, the processes (i)
and (ii) do not interfere in any order of perturbation series because, in Eq.~\eqref{me:as_1}, either the first line is
non-zero [process (i) is active] and the second line is zero
[process (ii) is inactive] or vice versa. Hence, we can write for the amplitude of scattering with $s=+$:
\begin{equation}
\label{sc_amplitude_T}
T_{\bm k+,\bm p -; \bm k' s_1, \bm p' s_1'}
= T_{\bm k\bm p,\bm k'\bm p'}\delta_{s_1,+}\delta_{s_1',-}
\pm 
(\bm k' \leftrightarrow \bm p', s_1 \leftrightarrow s_1'),
\end{equation}
where $T_{\bm k\bm p,\bm k'\bm p'} \equiv T_{\bm k+,\bm p -; \bm k' +, \bm p' -}$.
As a result we obtain the general expression for the collision integral
\begin{multline}
\label{St_via_T_general}
\text{St}[f_{\bm k+}] = {4\pi\over \hbar}\sum_{\bm p, \bm k', \bm p'}\delta(\varepsilon_k+\varepsilon_p-\varepsilon_{k'}-\varepsilon_{p'})\delta_{\bm k+\bm p,\bm k'+\bm p'}
\\ \times  \bigl[\abs{T_{\bm k\bm p,\bm k'\bm p'}}^2 f_{\bm k' +}f_{\bm p' -} (1\pm f_{\bm k+})(1\pm f_{\bm p-}) 
\\
- 
(\bm k, \bm p \leftrightarrow \bm k', \bm p')
\bigr],
\end{multline}
where the factor $4\pi = 2\times 2\pi$ comes from the account of the second term in Eq.~\eqref{sc_amplitude_T}.

In the first order in $V$, the scattering amplitude 
$T_{\bm k\bm p,\bm k'\bm p'}$ 
coincides with the matrix element:
\begin{equation}
\label{T_1_order}
V(\abs{\bm k - \bm k'})\braket{{\bm k},+}{{\bm k'},+}\braket{{\bm p},-}{{\bm p'},-} \equiv U_{\bm k\bm p,\bm k'\bm p'}.
\end{equation}
The collision integrals can be derived within the lowest Born approximation for the scattering matrix (we assume that the standard criteria for the smallness of the scattering potential are fulfilled, and for spin polarized fermions and bosons they are given, respectively, in Refs.~\cite{glazov04a} and \cite{glazov05a}, see also Refs.~\cite{ll1982,ll1982:II}). 
First, we present the expressions for the collision integrals responsible for the spin and valley current relaxation. Following Refs.~\cite{glazov04a,glazov05a,Glazov_2021b} we have from Eqs.~\eqref{St_via_T_general} and~\eqref{T_1_order}
\begin{multline}
\label{St:stand}
\text{St}_{\rm rel}[f_{\bm k+}] = {4\pi\over \hbar}\sum_{\bm p, \bm k', \bm p'}\delta(\varepsilon_k+\varepsilon_p-\varepsilon_{k'}-\varepsilon_{p'})\delta_{\bm k+\bm p,\bm k'+\bm p'}
\\ \times  \bigl[\abs{U_{\bm k\bm p,\bm k'\bm p'}}^2 f_{\bm k' +}f_{\bm p' -} (1\pm f_{\bm k+})(1\pm f_{\bm p-}) 
\\
- 
(\bm k, \bm p \leftrightarrow \bm k', \bm p')
\bigr].
\end{multline}

Here and in what follows we are interested in the spin current generation due to the skew scattering. Since the force itself due to the first term in the kinetic equation~\eqref{kinetic:gen} produces the same anisotropic contribution to the distribution function for both spin or valley states, we omit subscripts $s,s'$ in the distribution functions in Eq.~\eqref{St_via_T_general}. 
Our aim is to derive the asymmetric contribution to the collision integral responsible for the skew scattering. Similarly to the case of the impurity scattering, $|U_{\bm k \bm p, \bm k' \bm p'} |^2$ does not contain asymmetric part. The skew scattering probability can be obtained in the next to the first Born approximation~\cite{Sturman_1984}.
This means that in the standard collision integral~\eqref{St_via_T_general} one should take the squared modulus of the scattering amplitude, $\abs{T}^2$, in the third order in the scattering potential $U$, i.e. include the processes of scattering via intermediate states as well. 

The relevant processes are shown schematically in Fig.~\ref{fig:scatt3}: the diagram (a) demonstrates the first Born approximation, while the diagrams (b) and (c) show the processes where the transition $\bm k', \bm p' \to \bm k, \bm p$ takes place via intermediate states $\bm k''$ and $\bm p''$. Note that the process (b) can be interpreted as a result of consecutive interactions: Two particles occupying the initial states $\bm k'$ and  $\bm p'$ interact and scatter to the intermediate states $\bm k''$ and  $\bm p''$, respectively, then interact again and scatter to the final states $\bm k$ and $\bm p$. 
The process (c) requires interaction of the particle in any of initial state with the particle in the intermediate state, e.g., the particle in the initial $\bm k'$ state interacts with the particle in the intermediate $\bm p''$ state. As a result they scatter to $\bm k''$ and $\bm p$, respectively, then the particle that arrived in the intermediate state $\bm k''$ interacts with the one in the initial state $\bm p'$ and they scatter to $\bm k$ and $\bm p''$. Accordingly, this process is possible only in the presence of particles in intermediate states. The interference of the processes (a) and (b,c) results in the skew-scattering contributions we are looking for. We stress that the presence of the processes depicted in Fig.~\ref{fig:scatt3}(c), where the interactions with intermediate states play a role, does not allow one to use the standard form of the interparticle collision integral with the replacement of the Born scattering amplitude by the total scattering amplitude. To be specific, we assume that one particle with final, intermediate and initial wavevectors $\bm k$, $\bm k''$ and $\bm k'$ has $s=+$ and another particle scattered 
$\bm p' \to \bm p$ or $\bm p' \to \bm p'' \to \bm p$
has $s'=-$, and bear in mind that the final expressions for the scattering rates should contain corresponding permutations of the wavevectors $\bm k',\bm k''$ with $\bm p',\bm p''$. Under these assumptions 
we can write the Born perturbation series for $T_{\bm k\bm p,\bm k'\bm p'}$ only.  Importantly, in derivation of the $T_{\bm k\bm p, \bm k'\bm p'}$ a special care should be taken to account the occupancies of the intermediate states: As it is well known for the case of elastic scattering, only a proper account for all processes, including the ones where the intermediate state is occupied, results in the correct form of asymmetric terms in the collision integral~\cite{Sturman_1984}. In our case we need, therefore, to consider all possibilities for the intermediate states $\bm k'', \bm p''$: both can be empty or occupied, or one of these states can be empty and another one can be occupied.

Up to the second order in the interaction potential, the scattering amplitude has the form
\begin{multline}
T_{\bm k \bm p, \bm k' \bm p'} =U_{\bm k \bm p, \bm k' \bm p'}+
\sum_{\bm k'' \bm p''} \biggl\{ 
{U_{\bm k \bm p, \bm k'' \bm p''}U_{\bm k'' \bm p'', \bm k' \bm p'} \over \varepsilon_k+\varepsilon_p-\varepsilon_{k''}-\varepsilon_{p''}+i0}
\\ \times\delta_{\bm k+\bm p,\bm k''+\bm p''}
\qty[f_{\bm k''}f_{\bm p''} + (1\pm f_{\bm k''})(1\pm f_{\bm p''}) ]
\\ +
{U_{\bm k \bm p'', \bm k'' \bm p'}U_{\bm k'' \bm p, \bm k' \bm p''} \over \varepsilon_k+\varepsilon_{p''}-\varepsilon_{k''}-\varepsilon_{p'}+i0}
\\ \times\delta_{\bm k-\bm p',\bm k''-\bm p''}
\qty[f_{\bm k''}(1\pm f_{\bm p''}) + (1\pm f_{\bm k''})f_{\bm p''} ]
\biggr\}.
\end{multline}
Here the first term in the sum describes scattering via two occupied or two empty states while the second term describes scattering via one occupied and one empty state.

The skew scattering term in the lowest (3rd) 
order in $U$ is obtained from the interference of the 1st and 2nd order terms in $\abs{T_{\bm k \bm p, \bm k' \bm p'}}^2$  where the $\delta$-functions are taken from the energy denominators:
\begin{multline}
\label{T_quad}
\abs{T_{\bm k \bm p, \bm k' \bm p'}}^2_{\rm sk} =
4\pi\sum_{\bm k'' \bm p''}\Bigl\{ \text{Im}\qty[ U_{\bm k \bm p, \bm k' \bm p'}^* U_{\bm k \bm p, \bm k'' \bm p''}U_{\bm k'' \bm p'', \bm k' \bm p'}]
\\ \times
\delta(\varepsilon_k+\varepsilon_p-\varepsilon_{k''}-\varepsilon_{p''})\delta_{\bm k+\bm p,\bm k''+\bm p''}
\\ \times
\qty[f_{\bm k''}f_{\bm p''} + (1\pm f_{\bm k''})(1\pm f_{\bm p''}) ] 
\\ 
+\text{Im}\qty[ U_{\bm k \bm p, \bm k' \bm p'}^* U_{\bm k \bm p'', \bm k'' \bm p'}U_{\bm k'' \bm p, \bm k' \bm p''}]
\\ \times
\delta(\varepsilon_k+\varepsilon_{p''}-\varepsilon_{k''}-\varepsilon_{p'})\delta_{\bm k-\bm p',\bm k''-\bm p''}
\\ \times
\qty[f_{\bm k''}(1\pm f_{\bm p''}) + (1\pm f_{\bm k''})f_{\bm p''} ]
 \Bigr\}.
\end{multline}
Here the factor `4' accounts for the permutations $\bm k'' \leftrightarrow \bm p''$ in the intermediate states while the permutation $(\bm k'\leftrightarrow \bm p')$ is taken account in the prefactor in Eq.~\eqref{St_via_T_general}.

Within the minimal model the overlap of the
Bloch amplitudes with allowance for the spin-orbit coupling can be written as 
\begin{equation}
\label{ukuk1}
\langle {\bm k,s}|{\bm k',s}\rangle  = 1+ s\xi [\bm k \times \bm k']_z,
\end{equation}
where the real parameter $\xi$ describes the strength of the spin-orbit or valley-orbit coupling yielding~\cite{PhysRev.34.553,boguslawski,glazov2009,badalyan-2009,Glazov_2021b}
\begin{equation}
\label{U_2D_Dirac}
U_{\bm k \bm p, \bm k' \bm p'} = V(\abs{\bm k-\bm k'}) \qty(1+i\xi[\bm k \times \bm k' - \bm p \times \bm p']_z).
\end{equation}
In this work we disregard any spin-splitting of the energy spectrum, caused, e.g., by the Rashba effect. The electron-electron collision integral with allowance for the $\bm k$-linear terms in the spectrum is presented in Ref.~\cite{Mineev:2021vo} and it does not contain asymmetric contributions. 

Using Eq.~\eqref{U_2D_Dirac} and the general momentum-conservation condition $\bm k+\bm p=\bm k'+\bm p'$ we obtain that both the first and the second $\text{Im}[\ldots]$ in Eq.~\eqref{T_quad} are given by
\begin{equation}
\xi  V(\abs{\bm k-\bm k'}) V(\abs{\bm k-\bm k''}) V(\abs{\bm k''-\bm k'}) \mathcal S_z,
\end{equation}
where
\begin{equation}
\bm{\mathcal S}  = (\bm k'' +\bm p'' - \bm k - \bm p)\times (\bm k'- \bm k).
\end{equation}
We observe that, due to the momentum-conservation factor $\delta_{\bm k+\bm p,\bm k''+\bm p''}$, the first $\text{Im}[\ldots]$ in Eq.~\eqref{T_quad} does not contribute to the skew scattering probability. 
Thus, transitions described by the diagram in Fig.~\ref{fig:scatt3}(b)
do not contribute to the interparticle skew scattering rate in agreement with analysis in Ref.~\cite{Glazov_2021b}. In the case of fermions it means that the scattering processes where the two-particle intermediate states are empty or fully occupied play no role.
It already demonstrates significant difference of the asymmetric interparticle scattering as compared to the skew scattering by static impurities.

Nevertheless, the momentum conservation for the second $\text{Im}[\ldots]$ in Eq.~\eqref{T_quad} differs, therefore we obtain a nonzero result in general. We stress that an asymmetry of the quasiparticle collisions occurs at scattering processes described by Fig.~\ref{fig:scatt3}(c) where the particle in the initial state interacts with the particle in the intermediate state. For instance, for fermions it means that the skew scattering occurs only via such the two-particle states, where one single particle state is empty and another one is occupied. 
As we demonstrate below it results in unconventional density and temperature dependence of the SHE and VHE under interparticle collisions.

Assuming that the scattering is sufficiently short-range: $V(q) = V_0$, as it is the case in the gated two-dimensional electron systems and also for excitons, we finally obtain the asymmetric contribution to the interparticle collision integral:
\begin{multline}
\label{St:skew:fin}
\text{St}_{\rm sk} [f_{\bm k}] \\
= {16\pi^2\over \hbar} V_0^3\xi\sum_{\bm k' \bm p',\bm k'' \bm p'',\bm p} \delta(\varepsilon_k+\varepsilon_p-\varepsilon_{k'}-\varepsilon_{p'})\delta_{\bm k+\bm p,\bm k'+\bm p'} 
\\ \times
[(1\pm f_{\bm k})(1\pm f_{\bm p})f_{\bm k'}f_{\bm p'} + (\bm k, \bm p \leftrightarrow \bm k',\bm p')] 
\\ \times 
{\mathcal S}_z \delta(\varepsilon_k+\varepsilon_{p''}-\varepsilon_{k''}-\varepsilon_{p'})
\delta_{\bm k+\bm p'',\bm k''+\bm p'}
\\ \times 
\qty[f_{\bm k''}(1\pm f_{\bm p''}) + 
(\bm k'' \leftrightarrow \bm p'')
].
\end{multline}
Here we took into account that  the interchange $\bm k, \bm p \leftrightarrow \bm k',\bm p'$ changes the sign of $\bm{\mathcal S}$ while both energy and both momentum conservation laws in Eq.~\eqref{T_quad} are intact. Therefore the `out-scattering' term is added to the `in-scattering' one with the only modification in the occupation factors of the initial and final states. 

\begin{figure}[ht]
\includegraphics[width=0.75\linewidth]{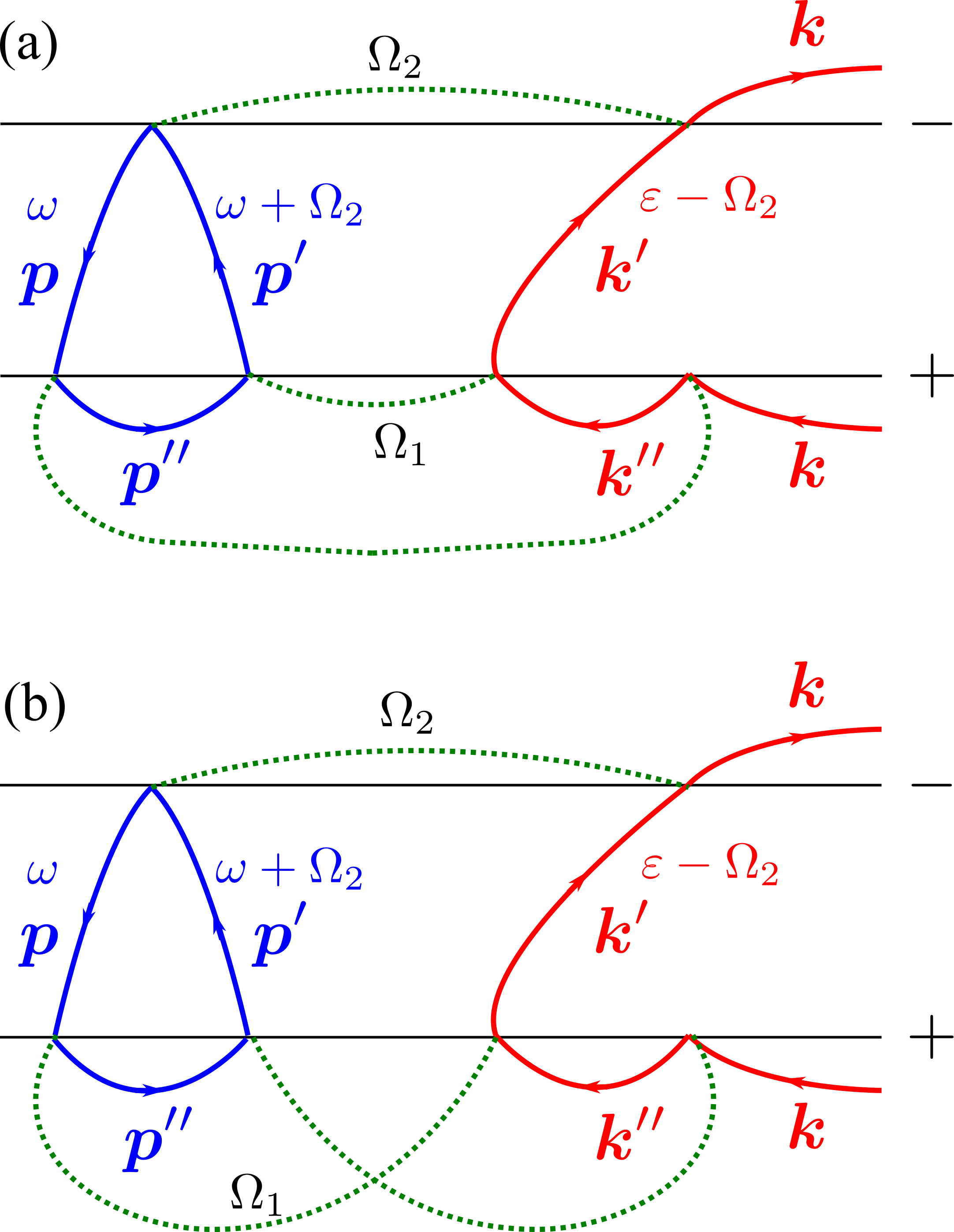}
\caption{Examples of the diagrams in the Keldysh technique describing the in-scattering processes. (a) Scattering via two empty or two occupied intermediate states, cf. Fig.~\ref{fig:scatt3}(b). (b)~Scattering with the intermediate states with one being empty and another being occupied, cf. Fig.~\ref{fig:scatt3}(c).} \label{fig:keldysh}
\end{figure}

Equation~\eqref{St:skew:fin} can be also derived using the Keldysh diagram technique. The in- and out-scattering terms can be recast as
\begin{equation}
\label{St:Keld}
\St[f_{\bm k}] = \mathrm i \Sigma^{-+}(1-f_{\bm k}) + \mathrm i \Sigma^{+-} f_{\bm k},
\end{equation}
where $\Sigma^{\alpha\beta}$ are the corresponding self-energies with $\alpha,\beta = \pm$ enumerating branches at the Keldysh contour, Fig.~\ref{fig:keldysh}. Two basic third order processes are shown in Fig.~\ref{fig:keldysh} with the diagram (a) and its flipped counterpart describing the transitions via the empty or occupied intermediate states [interaction lines do not cross, cf. Fig.~\ref{fig:scatt3}(b)], while the diagram (b) and its flipped counterpart describe the transitions where we take into account interaction of the particle in the initial and in one of the $\bm k''$, $\bm p''$ intermediate states [interaction lines cross, cf. Fig.~\ref{fig:scatt3}(c)]. Using the explicit expressions for the Green's functions and performing corresponding integrations (see Appendix~\ref{app:Keldysh}) we obtain the collision integral exactly in the same form as in Eq.~\eqref{St:skew:fin}.

The skew-scattering part of the collision integral $\text{St}_{\rm sk}$ is zero at the equilibrium distribution function $f_k^0$. This fact follows from the general arguments: for the spin-orbit coupling in the form of Eq.~\eqref{U_2D_Dirac} any equilibrium spin currents are forbidden. One can also demonstrate it explicitly making use of the fact that the equilibrium distribution functions depend on the energy only and using the energy and momentum conservation laws in Eq.~\eqref{St:skew:fin} and explicit form of $\mathcal S_z$, see Appendix~\ref{app:equilibr} for details.

\section{Spin and Valley Hall effects}
\label{sec:she}

Following Ref.~\cite{2020arXiv200405091G} we define the spin/valley current as
\begin{equation}
\label{Js:def}
\bm j_{\rm SHE/VHE} \equiv \bm j^{\rm s} = \frac{1}{2} \sum_{s,\bm k} s \frac{\hbar \bm k}{m} f_{\bm k,s}. 
\end{equation}
In the studied situation the electrons are unpolarized, thus it is sufficient to calculate the current for $s=+$ branch, while the current in the $s=-$ branch has the same magnitude and opposite direction. 
We now turn to the calculation of the SHE and VHE from the kinetic Eq.~\eqref{kinetic:gen} with the collision integrals~\eqref{St:skew:fin} and~\eqref{St:stand}. We treat the spin-orbit coupling as a small perturbation, accordingly, we solve Eq.~\eqref{kinetic:gen} by iterations.

We calculate the velocity generation rate:
\begin{equation}
\left< \dot{\bm V}\right> = \sum_{\bm k} {\hbar\bm k\over m} \text{St}_{\rm sk} [f_{\bm k}].
\end{equation}
Analysis shows (see Appendix~\ref{Kinematics}) that 
this generation rate is nonzero only if the following conditions are met  in Eq.~\eqref{St:skew:fin} for  $\text{St}_{\rm sk}$
\begin{equation}
\label{kinematics}
\bm p' = \bm k, \quad \bm k' = \bm p, \quad \bm p'' = \bm k''.
\end{equation}
In other words, the only relevant scattering processes are $\bm k+, \bm p- \rightarrow \bm k''+, \bm k''- \rightarrow \bm p+, \bm k-$.

Therefore we simplify the velocity generation rate
applying Eq.~\eqref{kinematics} everywhere except for the $\delta$-functions:
\begin{multline}
\label{eq:dot_V}
\left< \dot{\bm V}\right> = {64\pi^2\over m} V_0^3\xi
\\
\times
 \sum_{\bm k' \bm p',\bm k'' \bm p'', \bm k \bm p} \delta(\varepsilon_k+\varepsilon_p-\varepsilon_{k'}-\varepsilon_{p'})\delta_{\bm k+\bm p,\bm k'+\bm p'}
 \\ \times \delta(\varepsilon_k+\varepsilon_{p''}-\varepsilon_{k''}-\varepsilon_{p'})
\delta_{\bm k+\bm p'',\bm k''+\bm p'}
 \\ 
\times \bm k {\mathcal S}_z  (1\pm f_{\bm k})(1\pm f_{\bm p})f_{\bm k}f_{\bm p}
f_{\bm k''}(1\pm f_{\bm k''})
.
\end{multline}

 Since generation of the spin and valley Hall current is a nonequilibrium response to the external force $\bm F$, substituting the distributions $f_{\bm k}= f_k^0- \hbar (\bm v_{dr} \cdot \bm k)\dd f_k^0/\dd \varepsilon_k$ with the drift velocity $\bm v_{dr}=\bm F\tau_p/m$ and linearizing Eq.~\eqref{eq:dot_V} in $\bm v_{dr}$ we obtain 
\begin{widetext}
\begin{multline}
\label{dot_v}
\left< \dot{V}_y\right> = v_{dr,x} {16 \pi^2\hbar\over mT}V_0^3\xi 
\sum_{\bm k' \bm p',\bm k'' \bm p'',\bm k \bm p} \delta(\varepsilon_k+\varepsilon_p-\varepsilon_{k'}-\varepsilon_{p'})\delta_{\bm k+\bm p,\bm k'+\bm p'} 
(1\pm f_k^0)(1\pm f_p^0)f_{k}^0f_{p}^0f_{k''}^0(1\pm f_{ k''}^0) \\
\times \delta(\varepsilon_k+\varepsilon_{p''}-\varepsilon_{k''}-\varepsilon_{p'})
\delta_{\bm k+\bm p'',\bm k''+\bm p'}
\bm{\mathcal S} \cdot \qty{\qty[ \bm k \qty(1\pm 2  f_k^0)+\bm p \qty(1\pm 2  f_p^0) + \bm k'' \qty(1 \pm 2 f_{k''}^0)] \times (\bm k-\bm p) }
.
\end{multline}
\end{widetext}
Here we assume $\bm v_{dr} \parallel x$ and $\tau_p$ to be energy independent, accordingly we calculate the generation rate of the transverse to $\bm F$ velocity,
$\bm{\mathcal S} = (\bm k +\bm p - 2\bm k'')\times (\bm k- \bm p)$, and $T$ is the temperature in energy units.

In order to find the spin or valley current in the steady state one has to take into account its relaxation. If the dominant relaxation mechanism is the elastic scattering by impurities or quasielastic scattering by phonons, then 
\begin{subequations}
\label{Js:rel}
\begin{equation}
\label{Js:p}
j^s_y = \tau_p \left< \dot{V}_y\right> .
\end{equation}
However, we are mainly interested in the situation where the interparticle collisions dominate the relaxation. Conserving the total momentum of the colliding quasiparticles they result in the relaxation of the spin and valley currents. Generally, these collisions are inelastic and the solution of the corresponding kinetic equation is quite involved~\cite{glazov02,glazov04a}. To illustrate the effect and provide an estimate of its magnitude it is convenient to introduce a relaxation time approximation and present the current in the form similar to Eq.~\eqref{Js:p}
\begin{equation}
\label{Js:ee}
j^s_y = \tau_{\rm sc} \left< \dot{V}_y\right> 
\end{equation}
\end{subequations}
with $\tau_{\rm sc}$ being the corresponding spin current relaxation time, it is derived in Appendix~\ref{app:ee} and defined by Eq.~\eqref{tau:ee:def}. Note that by contrast to relatively slow relaxation of odd harmonics of spin-averaged distribution (or lack of relaxation in the case of the first harmonic)~\cite{LEDWITH2019167913}, the interpartcle collisions result in the efficient relaxation of both first and third angular harmonics of the spin and valley distrubtion~\cite{glazov04a}.

Still, the expressions for the SHE and VHE currents for arbitrary statistics of quasiparticles are quite involved. Below we consider the most illustrative and experimentally relevant limits of Fermi, Boltzmann and Bose gases.

\subsection{Fermi gas}\label{sec:fermi}

In the case of degenerate electrons, a dramatic suppression of the SHE and VHE at the electron-electron collisions occurs.
We recall that the electron-electron scattering time responsible for the spin current relaxation at $E_{\rm F} \gg T$ behaves as 
\begin{equation}
\label{tau:ee:deg}
\frac{1}{\tau_{\rm sc}} = C\frac{T^2}{E_F},
\end{equation}
where the constant $C$ weakly depends on the temperature~\cite{glazov04a}: The Pauli exclusion principle blocks electron-electron collisions at low temperatures. Inspecting general Eq.~\eqref{dot_v} we note that the factors $f^0_{k''}(1-f^0_{k''})$ should provide additional reduction of the phase space 
and one can expect $\left\langle \dot{V}_y \right\rangle \propto T^3$.

However the analysis shows that it is not the case, and $\left\langle \dot{V}_y \right\rangle$ is expected to behave as $T^4$. 
Indeed, to obtain a result in the lowest order in $T/E_F$ we can set all absolute values of the wavevectors equal to the Fermi wavevector $k=k'=k''=p=p'=p'' = k_{\rm F}$
everywhere except for the equilibrium occupations.
Then we come to the energy integrals ($q = k,p$ or $k''$)
\begin{equation}
\int_0^\infty \dd\varepsilon_q (1- f_q^0)f_{q}^0(1- 2  f_q^0)
\propto \int_0^\infty \dd\varepsilon_q \dv[2]{f_q^0}{\varepsilon_q} ,
\end{equation}
which are exponentially small at $E_{\rm F}\gg T$.

Hence, the non-zero result for the scattering rate can be obtained in the next order in $T/E_{\rm F}$ where one takes into account deviations of the absolute values of the wavevectors from $k_{\rm F}$. Consequently, we expect for degenerate electrons
\begin{equation}
j_y^s \sim N_1 v_{dr,x}\times \xi k_\text{F}^2 \times gV_0 \times \qty({T \over E_\text{F}})^2.
\end{equation}
As a result, the generation of the spin and valley currents due to the electron-electron scattering is suppressed. By contrast, interparticle scattering results in relatively efficient relaxation of the spin and valley currents. 

\subsection{Boltzmann gas}\label{sec:boltz}

For Boltzmann statistics we have for both bosons and fermions $f_k^0=N_1/(gT)\exp(-\varepsilon_k/T){ \ll 1}$ where $N_1$  and $g=m/( 2\pi\hbar^2)$ are the two-dimensional particle density and the density of states per one spin or valley. 

Calculations presented in Appendix~\ref{Append_Boltzmann} (see also Appendix~\ref{App:sc:Boltzmann}) yield
 the steady-state spin current value in the form
\begin{equation}
\label{SHE:Boltzmann}
j_y^s =
-{256\sqrt{3} \over \pi} N_1 v_{dr,x}\times \xi k_T^2 \times gV_0 \times {E_\text{F}\over T}.
\end{equation}
Here $N_1 v_{dr,x}$ is the particle current in one spin subband, $\xi k_T^2$ is the dimensionless spin-orbit coupling strength ($k_T=\sqrt{2mT}/\hbar$ is the thermal wavevector), $gV_0$ is the non-Born parameter, and the factor $E_\text{F}/T=4\pi N_1/k_T^2 \ll 1$ stems from occupations of intermediate states. In derivation we used the spin-current relaxation time for the non-degenerate case, Eq.~\eqref{tau:ee:Boltzmann}:
\begin{equation}
{1\over \tau_\text{sc}} = {2m\over \hbar^3}  V_0^2 N_1.
\end{equation}

It is instructive to compare the result~\eqref{SHE:Boltzmann} with the spin current induced by the scattering by the static short-range impurities with the same density $N_1$ in the absence of electron-electron collisions. In these situations the relaxation rates for the spin currents are about the same. From Eq.~(22) of Ref.~\cite{2020arXiv200405091G} we obtain 
\begin{equation}
\label{SHE:Boltzmann:imp}
j_y^s \sim  N_1 v_{dr,x}\times \xi k_T^2 \times gV_0,
\end{equation}
where the numerical coefficient on the order of unity is omitted. Comparing Eqs.~\eqref{SHE:Boltzmann} and \eqref{SHE:Boltzmann:imp} we observe that the spin and valley currents induced by the electron-electron collisions have additional smallness $\sim E_{\rm F}/T$ related, as discussed above, to the fact that the skew scattering requires 
the intermediate state $\bm k''$ to be occupied.
In the case of impurity scattering the intermediate state occupancies play no role thus parametrically enhancing the generation rate.
 However, while for reasonable electron densities $N_1 \sim 10^{11}\text{cm}^{-2}$, $m=0.1~m_0$ and at room temperature the small parameter $E_{\rm F}/T$ is about $0.1$,
the numerical prefactor in Eq.~\eqref{SHE:Boltzmann} $256\sqrt{3} / \pi \approx 141.1$ provides significant enhancement of the effect. Thus, in the state of the art samples the interparticle scattering effect on the spin current can be pronounced.

\subsection{Degenerate Bose gas}
\label{sec:bose}

While degeneracy in Fermi systems results in suppression of the interparticle scattering effects, in degenerate Bose gas stimulated scattering processed dominate. As a result, for degenerate bosons the occupancy of the intermediate states is expected to significantly enhance the skew scattering effect compared to the enhancement of the spin current relaxation rate.

To illustrate the enhancement we  observe that for the degenerate Bose gas,  the distribution function can be represented as $f_k^0=T/(\varepsilon_k-\mu) \gg 1$ with the chemical potential $\mu<0$, $\abs{\mu} \ll T$. For energies $\varepsilon_k - \mu \gtrsim T$ the distribution function exponential decays, such energy range is irrelevant for the following. Calculations presented in Appendix~\ref{Append:Bose} demonstrate  the exponentially large  steady-state value of the spin/valley current:
\begin{equation}
\label{SHE:Bose}
j_y^s =- \mathcal C \times
v_{dr,x} gT \times \xi k_T^2 \times  g V_0 \times\exp({2N_1 \over gT}).
\end{equation}
Here $gT$ is the characteristic concentration of particles where the gas becomes degenerate, consequently $2N_1/(gT) > 1$ for degenerate bosons, and the numerical coefficient calculated in Appendix~\ref{Append:Bose} is
 $\mathcal C\approx {1340.4}$. In deriving Eq.~\eqref{SHE:Bose} we used the value of the spin current relaxation rate~\eqref{sc_rel_rate_Bose}:
\begin{equation}
\frac{1}{\tau_{\rm sc}} =\mathcal C_\tau \times 2^{11}{g^2TV_0^2\over\hbar} \exp({N_1\over gT}),
\end{equation}
where $\mathcal C_\tau \approx 0.36$ is the numerical coefficient, for calculations see Appendix~\ref{App:sc:Bose}. 

Equation~\eqref{SHE:Bose} shows a possibility of a dominant role of the skew scattering  at low enough temperatures in Bose systems. However, the analysis presented above is valid for temperatures exceeding the temperature of the Berezinskii-Kosterlitz-Thouless transitions to the superfluid state. The analysis of the collective properties of bosons with the skew scattering processes included goes beyond the scope of the present work.

\section{Conclusion}\label{sec:concl}

In this work we have developed a theory of the spin and valley Hall effects in two-dimensional systems in the presence of frequent collisions between the quasiparticles. We have addressed both fermions, electrons in quantum wells or two-dimensional semiconductors, and bosons, e.g., excitons or exciton-polaritons. We have focused on the skew scattering effect where the colliding particles with the opposite spin or valley indices scatter in the opposite directions. The skew scattering results in the generation of a spin or valley current in the transverse direction to the particle current induced, e.g., by a real or synthetic electric field. The relaxation of the spin and valley current also assumed to occur due to the interparticle scattering.

We have derived the asymmetric -- skew -- contribution to the interparticle collision integral in the third order of perturbation theory in the interaction potential. We have demonstrated that occupancies of intermediate states play a crucial role in the effect: the skew scattering occurs in the processes where the particle in the initial state interacts with the particle in the intermediate state. For fermions, for example, the two-particle  intermediate states which are
empty of fully occupied play no role. It results in a specific behavior of the spin and valley current generation rates. In a non-degenerate gas the generation rate is proportional to the squared density of the particles, while in a degenerate Fermi gas the generation rate demonstrates $T^4$ temperature dependence. By contrast, for degenerate Bose gas this results in exponential enhancement of the spin and valley currents with increasing of particle concentration or decreasing the temperature.
Moreover, in the Boltzmann and Fermi gases, the parametrically small spin or valley current is additionally increased because of numerical factors on the order 
$10^3 \ldots 10^4$ that appear in the resulting expressions.

Hence, an asymmetry of the interparticle scattering results in deviations of a classical Fermi-liquid picture where in the processes of quasiparticle scattering occupancies of intermediate states play no role. By contrast, for bosons, the role of the skew scattering turns out to be significant when the Bose gas becomes degenerate. The analysis of the non-Fermi liquid effects for degenerate electrons and effects of asymmetry in interactions of bosons on collective phenomena is a separate problem for further studies.

\acknowledgments

We are grateful to I. S. Burmistrov for interest to this work and useful comments.
The financial support of the Russian Science Foundation (Project No. 22-12-00211) is acknowledged.
The work of L.~E.~G. was supported by the Foundation for the Advancement of Theoretical Physics and Mathematics ``BASIS''.

\appendix

\section{Derivation of the skew scattering integral within the Keldysh technique}\label{app:Keldysh}

It is instructive to derive the collision integral using the diagram technique for non-equilibrium processes. According to the general rules~\cite{ll10_eng,Arseev:2015eng}, the evolution of the distribution function is governed by the self-energies $\Sigma^{-+}$ and $\Sigma^{+-}$ which describe the in- and out-scattering processes, see Eq.~\eqref{St:Keld}. In the case of the skew scattering, the self-energies are given by the diagrams containing three interaction lines. Making use of the fact that, in the absence of retardation, the interaction line connect the same parts of the Keldysh contour, we  obtain that $\Sigma^{-+}$ is determined by the diagrams in Fig.~\ref{fig:keldysh} and their counterparts where the interaction occurs on the `$-$' line of the Keldysh contour (also, in agreement with the description of the interaction matrix element~\eqref{me:as_1}
the summation should be performed of the diagrams with permuted $\bm k'\leftrightarrow \bm p'$, $\bm k'' \leftrightarrow \bm p''$).
Using the explicit expressions for the Greens functions (top signs refer to the fermions and bottom signs  to the bosons)
\begin{subequations}
\begin{align}
&G^{-+}(\varepsilon, \bm k) = \pm 2\pi \mathrm i f_{\bm k} \delta(\varepsilon - \varepsilon_k),\\
&G^{+-}(\varepsilon, \bm k)= - 2\pi \mathrm i (1\mp f_{\bm k}) \delta(\varepsilon - \varepsilon_k),\\
&G^{--}(\varepsilon, \bm k)= \frac{1\mp f_{\bm k}}{\varepsilon - \varepsilon_k+\mathrm i 0} \pm \frac{f_{\bm k}}{\varepsilon - \varepsilon_k-\mathrm i 0},\\
&G^{++}(\varepsilon, \bm k) =  -[G^{--}(\varepsilon, \bm k)]^*,
\end{align}
\end{subequations}
we obtain the following contributions $\mathrm i \Sigma^{-+}$:
\begin{subequations}
\label{Sigma-+}
\begin{multline}
\label{sigma:a}
\mbox{Fig.~\ref{fig:keldysh}(a):}\quad \mathrm i \Sigma^{-+}_a(\varepsilon,\bm k)\\
=  4 \sum_{\bm k'\bm k'' \bm p''} U_{\bm k'\bm p',\bm k\bm p} U_{\bm k \bm p, \bm k'' \bm p''}U_{\bm k'' \bm p'', \bm k' \bm p'} \\
  \int \frac{d\Omega_1}{2\pi \mathrm i}\int \frac{d\Omega_2}{2\pi \mathrm i} G^{-+}(\varepsilon-\Omega_2,\bm k')G^{++}(\varepsilon-\Omega_2+\Omega_1,\bm k'') \\
  \times \int \frac{d\omega}{2\pi \mathrm i} G^{+-}(\omega,\bm p)G^{++}(\omega+\Omega_2-\Omega_1,\bm p'') G^{-+}(\omega+\Omega_2,\bm p'),
\end{multline}
\begin{multline}
\label{sigma:b}
\mbox{Fig.~\ref{fig:keldysh}(b):}\quad \mathrm i \Sigma^{-+}_b(\varepsilon,\bm k)\\
=  4 \sum_{\bm k'\bm k'' \bm p''} U_{\bm k'\bm p',\bm k\bm p} U_{\bm k \bm p, \bm k'' \bm p''}U_{\bm k'' \bm p'', \bm k' \bm p'} \\
  \int \frac{d\Omega_1}{2\pi \mathrm i}\int \frac{d\Omega_2}{2\pi \mathrm i} G^{-+}(\varepsilon-\Omega_1,\bm k')G^{++}(\varepsilon+\Omega_1-\Omega_2,\bm k'') \\
  \times \int \frac{d\omega}{2\pi \mathrm i} G^{+-}(\omega,\bm p)G^{++}(\omega+\Omega_1,\bm p'') G^{-+}(\omega+\Omega_2,\bm p'),
\end{multline}
Here the factor $4$ arises from the permutations of the initial and intermediate wavevectors. The inclusion of the diagrams with interaction on the `$-$' contour results in taking twice the real part of Eqs.~\eqref{Sigma-+}.
\end{subequations}

In Eq.~\eqref{sigma:a} we can integrate over $\omega$ using the $\delta$-function in $G^{+-}(\omega,\bm p)$ and over $\Omega_2$ using any of $\delta$-functions in $G^{-+}(\varepsilon-\Omega_2, \bm k')$ or $G^{-+}(\omega+\Omega_2,\bm p')$. As a result, we obtain
\begin{multline}
\label{sigma:a:1}
\mathrm i \Sigma_a^{-+}(\varepsilon,\bm k) \propto \delta(\varepsilon+\varepsilon_{\bm p} - \varepsilon_{\bm k'} - \varepsilon_{\bm p'})(1-f_{\bm p})f_{\bm k'} f_{\bm p'}\\
\times  \int \frac{d\Omega_1}{2\pi \mathrm i}G^{++}(\varepsilon_{\bm k'}+\Omega_1,\bm k'')G^{++}(\varepsilon_{\bm p'}-\Omega_1,\bm p'').
\end{multline} 
To be non-zero, the poles in the integral over $\Omega_1$ should have different imaginary parts. Hence, in both $G^{++}$ Green's functions only the terms with similar occupation factors $\propto f_{\bm k''}f_{\bm p''}$ and $(1\mp f_{\bm k''})(1\mp f_{\bm p''})$ do not vanish. Integrating using the residue theorem and adding the conjugate term we arrive at the contribution to the in-scattering rate given by the first three lines in Eq.~\eqref{T_quad} of the main text.

Analogous calculation of $\Sigma_b^{-+}$ in \eqref{sigma:b} yields
\begin{multline}
\label{sigma:b:1}
\mathrm i \Sigma_{b}^{-+}(\varepsilon,\bm k) \propto \delta(\varepsilon+\varepsilon_{\bm p} - \varepsilon_{\bm k'} - \varepsilon_{\bm p'})(1-f_{\bm p})f_{\bm k'} f_{\bm p'}\\
\times  \int \frac{d\Omega_1}{2\pi \mathrm i}G^{++}(\varepsilon_{\bm k'}+\Omega_1,\bm k'')G^{++}(\varepsilon_{\bm p}+\Omega_1,\bm p'').
\end{multline} 
The integral over $\Omega_1$ in this case does not vanish only if unlike terms in the product of Greens functions is taken: either $f_{\bm k''}(1\mp f_{\bm p''})$ or $f_{\bm p''}(1\mp f_{\bm k''})$. Corresponding integration results in the contribution to the in-scattering rate given by the last three lines in Eq.~\eqref{T_quad} of the main text. Summing Eqs.~\eqref{sigma:a:1} and \eqref{sigma:b:1} we arrive at the in-scattering term of Eq.~\eqref{St:skew:fin} of the main text. The evaluation of the out-scattering terms is similar. Total collision integral fully agrees with 
\eqref{St:skew:fin} of the main text.

\section{Absence of the spin current in equilibrium}
\label{app:equilibr}

Let us take an isotropic distribution function $f_{k}$ (i.e., in the absence of external field) and consider the angular Fourier harmonics of the skew scattering integral, Eq.~\eqref{St:skew:fin}, 
\begin{align}
S_n = \sum_{\bm k} g_k \sin{(n\varphi_{\bm k})}\text{St}_{\rm sk} [f_k], \nonumber \\
 C_n = \sum_{\bm k} g_k \sin{(n\varphi_{\bm k})}\text{St}_{\rm sk} [f_k]. \nonumber
\end{align}
Here $g_k$ is an arbitrary function of electron energy (that vanishes at $k\to \infty$), and $n=0,1,\ldots$ Performing the change of the variables $(k_x,k_y)\to(-k_x,k_y)$, $(p_x,p_y)\to(-p_x,p_y)$ (and similar for $\bm k', \bm p', \bm k'', \bm p''$) which corresponds to the reflection in $(yz)$ plane we see that ${\mathcal S}_z$ changes its sign while all other multipliers are intact. As a result, $S_n\equiv 0$. Similar analysis with the reflection in the $(xz)$ plane shows that $C_n \equiv 0$. As a result,  $\text{St}_{\rm sk} [f_k]$ vanishes for any angular-independent distribution function $f_k$. In particular, it is zero for equilibrium distribution function. As a result, no spin and valley current can flow in equilibrium conditions for the spin-orbit interaction in the form of Eq.~\eqref{ukuk1}.

\section{Kinematics of the skew scattering}
\label{Kinematics}

In order to prove the relations~\eqref{kinematics}
we introduce
\begin{equation}
\label{Q_kappa_def}
\bm \varkappa=\bm k-\bm p, \quad \bm \varkappa'=\bm k'-\bm p', \quad 
\bm Q = \bm k+\bm p - (\bm k''+\bm p'').
\end{equation}

Then 
we have
\begin{multline}
\bm{\mathcal S} \delta(\varepsilon_k+\varepsilon_p-\varepsilon_{k'}-\varepsilon_{p'})\delta(\varepsilon_k+\varepsilon_{p''}-\varepsilon_{k''}-\varepsilon_{p'})
\\
={\bm Q\times(\bm \varkappa - \bm \varkappa')\over 2}
\qty({4m\over\hbar^2})^2 \delta(\varkappa^2-\varkappa'^2) 
\delta[\bm Q\cdot(\bm \varkappa {+} \bm \varkappa')].
\end{multline}
The second $\delta$-function implies one of 3 possibilities: i)~$\bm Q=\bm 0$, ii)~$\bm Q \perp (\bm \varkappa {+} \bm \varkappa')$, and iii)~$(\bm \varkappa {+} \bm \varkappa')=0$.
In the cases i) and~ii) 
$\bm{\mathcal S}$ vanishes, therefore we obtain a nonzero result in the case~iii) only. This yields the relations~\eqref{kinematics}.

Taking in Eq.~\eqref{dot_v} $\bm \varkappa' =-\bm \varkappa$ everywhere except for the $\delta$-function, 
we perform summation over $\bm \varkappa'$:
\begin{multline}
\sum_{\bm \varkappa'}\delta(\varkappa^2-\varkappa'^2) \delta[\bm Q \cdot(\bm \varkappa {+} \bm \varkappa')]
\\= {1\over 4\pi} \left<\delta[Q\varkappa (\cos \varphi+\cos\varphi')] \right>_{\varphi'}
= {1\over 8\pi^2 Q\varkappa \abs{\sin\varphi}} ,
\end{multline}
where $\varphi$ is an angle between $\bm Q$ and $\bm \varkappa$. Here only one root $\varphi'=\varphi + \pi$ was used because the other, $\varphi'=\pi-\varphi$, corresponds to $\bm Q \perp (\bm \varkappa+\bm \varkappa')$ which yields zero, see above.
Then we have:
\begin{multline}
\label{dot_V_1}
\left< \dot{V}_y\right> =v_{dr,x} {16 m \over T \hbar^3}  V_0^3\xi  \sum_{\bm K, \bm \varkappa,\bm Q} 
{[\bm Q \times \bm \varkappa]_z\over \abs{\bm Q \times \bm \varkappa}} 
\\ \times
(1\pm f_k^0)(1\pm f_p^0)f_{k}^0f_{p}^0f_{k''}^0(1\pm f_{ k''}^0)
\\ \times
\qty[\bm K \times \bm \varkappa (3 \pm 2 f_k^0 \pm 2 f_p^0 \pm 2 f_{k''}^0)  \mp 2  \bm Q \times \bm \varkappa f_{k''}^0]_z
.
\end{multline}
Here we introduced $\bm K=\bm k + \bm p$.

\section{Relaxation of spin and valley currents due to interparticle collisions}\label{app:ee}

It is convenient to introduce the pseudospin component $$s_{\bm k} = \frac{1}{2} \sum_{s} s f_{\bm k,s}$$ and obtain from the collision integrals~\eqref{St:stand} the contribution responsible for the relaxation of the pseudospin [cf. Refs.~\cite{glazov04a,glazov05a}]:
\begin{multline}
\label{St:sz:rel}
\St{[s_{\bm k}]} =  -\frac{4\pi V_0^2}{\hbar} \sum_{\bm k'\bm p\bm p'} \delta_{\bm k+\bm p, \bm k'+\bm p'} \delta(\varepsilon_k + \varepsilon_p - \varepsilon_{k'} - \varepsilon_{p'}) \\
\times \left[s_{\bm k} F(\bm p; \bm k', \bm p') - s_{\bm k'} F(\bm p'; \bm k, \bm p)\right],
\end{multline}
where
\[
F(\bm p; \bm k', \bm p') 
= f_{\bm p}(1\pm f_{\bm k'} \pm f_{\bm p'}) \mp f_{\bm p'} f_{\bm k'}.
\]

To determine the relaxation time $\tau_{\rm sc}$ we take $s_{\bm k}$ in the form $s_{\bm k} = A k_x f_{k}^{0}(1\pm f_{k}^{0})$ which corresponds to a variational solution of kinetic equation ($A$ is a variational parameter) with the generation rate $G k_x f_{k}^{0}(1\pm f_{k}^{0})$. The relaxation time is determined from the relation $A= \tau_{\rm sc} G$. To find $A$ we substitute variational solution into the kinetic equation, multiply it by $\cos{\varphi_{\bm k}}$ where $\varphi_{\bm k}$ is the angle between the $x$-axis and $\bm k$ and sum over $\bm k$ with the result
\begin{equation}
\label{tau:ee:def}
\frac{1}{\tau_{\rm sc}} = - \frac{\sum_{\bm k} \cos{\varphi}_{\bm k} \St{[k_x f_{k}^{0}(1\pm f_{k}^{0}) ]}}{\frac{1}{2} \sum_{\bm k} k f_{k}^{0}(1\pm f_{k}^{0})}.
\end{equation}

Substitution of Eq.~\eqref{St:sz:rel} yields
\begin{multline}
\label{tau_sc_general}
\frac{1}{\tau_{\rm sc}} =  \frac{4\pi V_0^2/\hbar}{ \sum_{\bm k} k f_{k}^{0}(1\pm f_{k}^{0})} 
\\
\times
\sum\limits_{\bm k \bm p \bm k' \bm p'} \delta_{\bm k+\bm p, \bm k'+\bm p'} \delta(\varepsilon_k + \varepsilon_p - \varepsilon_{k'} - \varepsilon_{p'})
\\
\times
\qty(k-\bm k \cdot \bm k'/k) f_{k}^{0}(1\pm f_{k}^{0})F(\bm p; \bm k', \bm p') ,
\end{multline} 
where we made substitutions $\cos{\varphi}_{\bm k}k_x \to {k/ 2}$,  $\cos{\varphi}_{\bm k} k_x' \to {\bm k \cdot \bm k'/ (2k)}$ taking into account the isotropy in the $xy$ plane.
Then using the notations $\bm \varkappa$, $\bm \varkappa'$ introduced in Eq.~\eqref{Q_kappa_def} we obtain
\begin{equation}
\label{tau_sc_general1}
{1\over \tau_\text{sc}}={2mV_0^2\over \hbar^3}
{\sum\limits_{\bm k,\bm \varkappa, \bm \varkappa'} 
\left< {\bm k \cdot (\bm \varkappa - \bm \varkappa')\over k}
f_{k}^{0}(1\pm f_{k}^{0})F(\bm p; \bm k', \bm p') \right>
\over \sum\limits_{\bm k}  k f_{k}^{0}(1\pm f_{k}^{0})}.
\end{equation}
Here angular brackets mean averaging over directions of the vector $\bm \varkappa'$ at $\varkappa'=\varkappa$, and $\bm p = \bm k - \bm \varkappa$, $\bm k'=\bm k+(\bm \varkappa' -\bm \varkappa)/2$, $\bm p'=\bm k-(\bm \varkappa' +\bm \varkappa)/2$.

\subsection{Relaxation rate at Boltzmann statistics}
\label{App:sc:Boltzmann}

For non-degenerate particles (Boltzmann statistics) we take in Eq.~\eqref{tau_sc_general1} $f^0_k = (N_1/gT)\exp(-\varepsilon_k/T) \ll 1$.
Then
we obtain
\begin{equation}
{1\over \tau_\text{sc}}={2mV_0^2\over \hbar^3}
{\sum\limits_{\bm k,\bm \varkappa} 
{\bm k \cdot \bm \varkappa \over k}
f_{k}^{0}f_{\bm k- \bm \varkappa}^{0} 
\over \sum\limits_{\bm k}  k f_{k}^{0}}.
\end{equation}

Performing angular averaging with help of $\left<\cos{\theta}\exp(a\cos{\theta}) \right>_\theta = I_1(a)$,
we get
\begin{equation}
{1\over \tau_\text{sc}} =
{4\pi V_0^2N_1\over \hbar T k_T^3/(8\sqrt{\pi})} 
\sum\limits_{\bm k,\bm \varkappa} 
\varkappa I_1\qty({2k\varkappa\over k_T^2})
\exp(-{2k^2+\varkappa^2 \over k_T^2}).
\end{equation}
Then integrating over absolute values of $k$ and $\varkappa$ we 
obtain
\begin{equation}
\label{tau:ee:Boltzmann}
{1\over \tau_\text{sc}}= {2 m\over \hbar^3}  V_0^2 N_1.
\end{equation}

\subsection{Relaxation rate at Bose-Einstein statistics}
\label{App:sc:Bose}

For degenerate Bose particles
we take in Eq.~\eqref{tau_sc_general1} 
\begin{equation}
\label{bose:deg}
f^0_k = T/(\varepsilon_k+\abs{\mu}) \gg 1,
\end{equation} where $\mu<0$ is related to the particle density with a fixed spin/ in a given valley via $N_1=gT\ln(T/\abs{\mu})$. 
Then we obtain
\begin{equation}
\frac{1}{\tau_{\rm sc}} =\frac{2^9 m V_0^2 g^2 T^4}{\hbar^3\mu^2} \sqrt{2m\abs{\mu}\over \hbar^2}
{1\over \sum_{\bm k} k \qty(f_{k}^0)^2} \mathcal C_\tau,
\end{equation}
where
\begin{multline}
\label{Ctau}
\mathcal C_\tau=
\\
\int\limits_0^\infty \dd X \int\limits_0^\infty \dd Y
 \Biggl<
{Y(1-\cos\varphi)+\sqrt{XY}[\cos(\theta-\varphi)-\cos\varphi]\over \sqrt{X + Y + 2\sqrt{XY}\cos\theta}}
 \\ \times
{1\over (X + Y+1)^2 - 4XY\cos^2(\theta-\varphi)}
\\
\times
 {1\over (X+Y+1)^2-4XY\cos^2\theta}
\Biggr>_{\varphi,\theta}
.
\end{multline}
Here we used $X=\varepsilon_{\bm k+ \bm p}/\abs{\mu}$, $Y=\varepsilon_{\varkappa}/\abs{\mu}$. Note that while the simplified form of the distribution function~\eqref{bose:deg} is valid for $\varepsilon_k \ll T$, the integrals in Eq.~\eqref{Ctau} converge at $X,Y\to \infty$, thus we can extend the integration up to $+\infty$.

Since 
\begin{equation}
 \sum_{\bm k} k \qty(f_{k}^0)^2 =  {\pi gT^2\over 2\abs{\mu}} \sqrt{2m\abs{\mu}\over \hbar^2},
\end{equation}
we obtain
\begin{equation}
\label{sc_rel_rate_Bose}
\frac{1}{\tau_{\rm sc}} =2^{11}{(gT)^2V_0^2\over\hbar\abs{\mu}} \mathcal C_\tau.
\end{equation}

For numerical evaluation of the 4-fold integral~\eqref{Ctau}  it is convenient to change the variables formally introducing $R$ and $\alpha$ as
$X=R\cos{\alpha}$, $ Y = R\sin{\alpha}$.
Correspondingly, the integration is carried out over a sector in the plane $0\leqslant \alpha \leqslant \pi/2$, and $R\in [0,\infty)$. Numerical calculation shows that
\begin{equation}
\label{Ctau:n}
\mathcal C_\tau \approx 0.36.
\end{equation}

\begin{widetext}
\section{Spin and valley current 
generation rate 
at Boltzmann statistics}
\label{Append_Boltzmann}

At Boltzmann statistics we have

\begin{equation}
f_{k'}^0f_{p'}^0f_{k''}^0 = f_{ {\abs{\bm K - \bm \varkappa}\over 2}}^0f_{ {\abs{\bm K + \bm \varkappa}\over 2}}^0f_{ {\abs{\bm K - \bm Q} \over 2}}^0
= \qty({N_1\over gT})^3 \exp[-{\hbar^2\over 4mT}\qty({3\over2}K^2 + \varkappa^2 + {1\over 2}Q^2 - \bm K \cdot \bm Q)],
\end{equation}
and obtain from Eq.~\eqref{dot_V_1}:
\begin{equation}
\left< \dot{V}_y\right> =v_{dr,x} {16 m \over T \hbar^3}  V_0^3\xi \qty({N_1\over gT})^3
\sum_{\bm K, \bm \varkappa,\bm Q} 
\exp[-{\hbar^2\over 4mT}\qty({3\over2}K^2 + \varkappa^2 + {1\over 2}Q^2 -  K Q\cos\theta)] {\sin\varphi\over \abs{\sin\varphi}}
 \qty[3 K\sin(\theta-\varphi)- Q\sin\varphi] \varkappa
.
\end{equation}
Here $\theta$ is an angle between $\bm K$ and  $\bm Q$, 
and $\varphi$ is the angle between $\bm Q$ and $\bm \varkappa$.

Averaging over 
$\theta$
yields:
\begin{equation}
\left< \exp(a\cos\theta)\right>_\theta = I_0(a), \qquad \left< \exp(a\cos\theta)\sin(\theta-\varphi)\right>_\theta =-\sin\varphi I_1(a),
\end{equation}
and we get
\begin{equation}
\left< \dot{V}_y\right> =-v_{dr,x} {{16} m \over T \hbar^3}  V_0^3\xi \qty({N_1\over gT})^3 \sum_{\bm K, \bm \varkappa,\bm Q} 
\exp[-{\hbar^2\over 4mT}\qty({3\over2}K^2 + \varkappa^2 + {1\over 2}Q^2 )] {\sin^2\varphi\over \abs{\sin\varphi}}
 \qty[3 K I_1\qty({\hbar^2K Q\over 4mT})+ QI_0\qty({\hbar^2K Q\over 4mT})] \varkappa
.
\end{equation}
Averaging over $\varphi$ and summation over $\bm \varkappa$ yield:
\begin{equation}
\left< {\sin^2\varphi\over \abs{\sin\varphi}}\right>_\varphi = {2\over \pi},
\qquad
\sum_{\bm \varkappa}\varkappa\exp(-{\hbar^2\varkappa^2\over 4mT}) ={1\over \sqrt{\pi}} \qty({mT\over \pi \hbar^2})^{3/2},
\end{equation}
therefore we get
\begin{equation}
\left< \dot{V}_y\right> =-v_{dr,x} {{32} m \over T \hbar^3\sqrt{\pi}}  V_0^3\xi\qty({N_1\over gT})^3 \qty({mT\over \pi \hbar^2})^{3/2}
\sum_{\bm K,\bm Q} 
\exp[-{\hbar^2\over 4mT}\qty({3\over2}K^2 + {1\over 2}Q^2 )] 
 \qty[3 K I_1\qty({\hbar^2K Q\over 4mT})+ QI_0\qty({\hbar^2K Q\over 4mT})] 
.
\end{equation}
Calculations show:
\begin{multline}
\sum_{\bm K,\bm Q} 
\exp[-{\hbar^2\over 4mT}\qty({3\over2}K^2 + {1\over 2}Q^2 )] 
3 K I_1\qty({\hbar^2K Q\over 4mT})= 
\sum_{\bm K,\bm Q} 
\exp[-{\hbar^2\over 4mT}\qty({3\over2}K^2 + {1\over 2}Q^2 )]  QI_0\qty({\hbar^2K Q\over 4mT}) \\
= \qty({mT\over \hbar^2})^{5/2} {2\sqrt{3}\over \pi^{3/2}}.
\end{multline}
Therefore we finally get for the generation rate:
\begin{equation}
\left< \dot{V}_y\right> =-N_1 v_{dr,x} {256\sqrt{3} \over \pi }  gV_0 \xi k_T^2 {E_{\rm F}\over T} \times {1\over \tau_{\rm sc}}.
\end{equation}

\section{Calculation of the spin current 
at Bose statistics}
\label{Append:Bose}

At $f^0_k = T/(\varepsilon_k+\abs{\mu}) \gg 1$ we obtain from Eq.~\eqref{dot_V_1}:
\begin{equation}
\left< \dot{V}_y\right>=-v_{dr,x} 2^{20} \pi g^4  V_0^3\xi k_T^2 {T^5\over \hbar\abs{\mu}^3}   \mathcal C_g
,
\end{equation}
where
\begin{multline}
\mathcal C_g = \int\limits_0^\infty \dd X_K \int\limits_0^\infty \dd X_Q \int\limits_0^\infty \dd X_\varkappa
 \Biggl< 
 {\abs{\sin\varphi} \sqrt{X_\varkappa} \over \qty[\qty(X_K + X_\varkappa +1)^2-4X_K X_\varkappa\cos^2(\theta-\varphi))^2]^2\qty(X_K + X_Q - 2\sqrt{X_KX_Q}\cos\theta+1)^2}
\\ \times
\qty[{2\qty(X_K + X_\varkappa +1) \sqrt{X_K}\cos\theta \over \qty(X_K + X_\varkappa +1)^2- 4X_KX_\varkappa\cos^2(\theta-\varphi)} + {\sqrt{X_K}\cos\theta+\sqrt{X_Q}  \over X_K + X_Q - 2\sqrt{X_KX_Q}\cos\theta+1} ]
\Biggr>_{\varphi,\theta}
.
\end{multline}
Here we introduced dimensionless energies $X_q = \varepsilon_{q}/\abs{\mu}$ ($q=K,Q,\varkappa$), and, as before, the upper limits for integration were extended to $+\infty$ due to the convergence of integrals.

Numerical calculation yields $\mathcal C_g \approx 0.3$.

Using Eq.~\eqref{sc_rel_rate_Bose} we can rewrite the velocity generation rate in the form
\begin{equation}
\left< \dot{V}_y\right> 
 = - v_{dr,x} gT \xi k_T^2  g V_0 {T^2\over \abs{\mu}^2} \times { \mathcal C\over \tau_{\rm sc}},
\end{equation}
where $\mathcal C = { 2^{9} \pi    \mathcal C_g 
 / 
\mathcal C_\tau} \approx {1340.4}$.
\end{widetext}

\bibliography{eeskew}

\end{document}